\documentclass[twocolumn]{autart} 

\usepackage{graphicx}      
\usepackage[square,sort,comma,numbers]{natbib}
\usepackage{color}
\usepackage{comment}
\usepackage{amsmath} 
\usepackage{amssymb}  
\usepackage{subcaption}

\newtheorem{proposition}{Proposition}
\newtheorem{lemma}{Lemma}
\newtheorem{theorem}{Theorem}

\newtheorem{definition}{Definition}

\newtheorem{remark}{Remark}
\newtheorem{example}{Example}

\begin{document}
\begin{frontmatter}

\title{Networked Supervisor Synthesis Against Lossy Channels with Bounded Network Delays as Non-Networked Synthesis\thanksref{footnoteinfo}} 

\thanks[footnoteinfo]{The research of the project was supported by Ministry of Education, Singapore, under grant AcRF TIER 1-2018-T1-001-245 (RG 91/18) and supported by the funding from Singapore National Research Foundation via Delta-NTU Corporate Lab Program (DELTA-NTU CORP
LAB-SMA-RP2 SU RONG M4061925.043). Liyong Lin and Yuting Zhu contribute equally.}

\author[First]{Liyong Lin*},
\author[First]{Yuting Zhu*},
\author[First]{Ruochen Tai},
\author[Second]{Simon Ware}, 
\author[First]{Rong Su} 


\address[First]{School of Electrical and Electronic Engineering, Nanyang Technological University, Singapore (e-mail: liyong.lin@ntu.edu.sg,
yuting002@e.ntu.edu.sg,
ruochen001@e.ntu.edu.sg, rsu@ntu.edu.sg).}
\address[Second]{Grab Holdings Inc., Singapore (email: simianware@gmail.com).}

\begin{abstract}                
In this work, we study the problem of supervisory control of networked discrete event systems. We consider lossy communication channels with bounded network delays, for both the control channel and the observation channel. By a model transformation, we transform the  networked supervisor synthesis problem into the classical (non-networked) supervisor synthesis problem (for non-deterministic plants), such that the existing supervisor synthesis tools can be used for synthesizing  networked supervisors. In particular, we can use the (state-based) normality property for the synthesis of the supremal networked supervisors, whose  existence is guaranteed by construction due to our consideration of command non-deterministic supervisors. The effectiveness of our approach is illustrated on a mini-guideway example that is adapted from the literature, for which the supremal networked supervisor has been synthesized in the synthesis tools SuSyNA and TCT.  

\end{abstract}


\end{frontmatter}



%

\section{Introduction}
%
%
%
%
In networked discrete event systems, the supervisor and the plant are (remotely) connected through  communication channels which are often imperfect. The observation messages from the plant and the control messages issued by the supervisor may  experience non-negligible network delays and may even get lost. Maintaining the safe operation of a supervisory control system against imperfect communication channels seems to be conceptually more difficult than the non-networked counterpart, especially when the communication channels are non-FIFO. 


Many papers have been devoted to the verification and synthesis of networked supervisors, considering network delays and/or communication losses. In  \cite{balemi1992supervision} and \cite{balemi1994input}, the solution for the networked supervisor synthesis problem, with bounded communication delays and under full observation, relies on the existence of a controllable,  delay-insensitive language. Later,  \cite{park2002robust}, \cite{park2006delay}, \cite{park2007supervisory} address the networked supervisor synthesis problem under the  assumption that  all the controllable events are observable and each controllable event can be executed only when a control message that enables this event  is received. \cite{lin2014control} introduces two observation mappings that model the effects of observation delays and observation losses,  respectively. Based on the two observation mappings, a necessary and sufficient condition for the existence of a networked supervisor is provided based on newly provided definitions of network observability and network controllablity. \cite{Shu2015} considers the networked supervisor synthesis problem, with bounded network delays in both the control channel and the observation channel, by specifying the behaviour of the supervised system with a minimum required language and a maximum admissible language. 
Some other recent works that also address the networked supervisor synthesis problem (in the centralized setup) include the works of~\cite{Lin:17},~\cite{Shu:17},~\cite{CDC:17},~\cite{rashidinejad2018supervisory} and~\cite{rashidinejad2019}.  
There are also several papers that have considered the   decentralized supervisory control  \cite{shu2014decentralized},  \cite{tripakis2004decentralized},  \cite{park2007decentralized} and the modular control~\cite{Komenda:16} of networked systems. The works of~\cite{LinWang19} and~\cite{Alves2019ACC} address the problem of state estimation for networked systems.

In this paper, we consider the problem of networked supervisor synthesis against lossy channels, with bounded network delays in both the control channel and the observation channel. This paper explicitly models the (FIFO or non-FIFO) observation channel and the control channel as automata over some lifted alphabets. We  consider both  channels to have an  infinite capacity but impose a natural restriction on the networked supervisors to ensure the regularity of the networked closed-loop systems, certified by a (non-linear) inductive invariant.  In particular, the inductive invariant establishes the tight upper bound on the effective channel capacity\footnote{In contrast,  the tight upper bound on the effective channel capacity for the observation channel is straightforward.} (for the control channel). The reason for us to consider channels with an infinite capacity is to ensure the  upper bound of the effective channel capacity (for the control channel), after we impose the restriction on the networked supervisors, is obtained as a by-product of the analysis instead of being imposed a prior (as done in~\cite{rashidinejad2019} and~\cite{zhusupervisor}); in particular, we are interested in tight upper bounds to minimize the state sizes of the channel automata that need to be constructed in the worst case.  

The idea to model communication channels as automata is not quite new (see, for example~\cite{rashidinejad2018supervisory},~\cite{rashidinejad2019},~\cite{tripakis2004decentralized},~\cite{LinWang19},~\cite{Alves2019ACC}).  ~\cite{tripakis2004decentralized} only models the lossless FIFO observation channels, since the control channel of~\cite{tripakis2004decentralized} is  assumed to  involve no delay.~\cite{LinWang19}  and~\cite{Alves2019ACC} only consider and model the observation channels, since only the networked state estimation problem, instead of the networked supervisor synthesis problem, is considered.
This paper is much more closely related to~\cite{rashidinejad2018supervisory}, where  both the control channel and the observation channel are   modelled as automata, but with some important differences. 
 
 In this work, each control message in the control channel encodes a control command,  which is a set of (enabled) controllable events,  following the Ramadge-Wonham supervisory control framework~\cite{wonham2015supervisory}. In contrast, each control message in the control channel of~\cite{rashidinejad2018supervisory} encodes an (enabled) controllable event. The use of control messages that encode control commands entails the modelling of a command execution automaton  that ``transduces" control commands to events executed in the plant. 
In this work, we  consider it is  possible that  multiple observation messages (respectively, multiple control messages) are received simultaneously\footnote{Simultaneous receptions of control messages refers to the scenario where multiple control messages are received by the plant  (consecutively), within one time step, before any plant event is executed. Similarly, simultaneous receptions of observation messages refers to the scenario where multiple observation messages are received by the networked supervisor  (consecutively), before any control message is issued. This situation is exacerbated by our assumption of bounded network delays, instead of fixed network delays~\cite{rashidinejad2018supervisory}. } by the networked supervisor (respectively, by the plant). We shall consider the practical scenario where both the networked supervisor and the plant are always ready to react to and process these received messages. We use asynchronous event interleaving, captured by the standard synchronous product operation~\cite{wonham2015supervisory}, to simulate\footnote{Note that different modules in a networked discrete-event system are  synchronized  through the sending and receiving of  messages. The  plant and the command execution automaton  are synchronized through shared plant events, and thus they are together treated as an independent  module, referred to as an (augmented) plant.  More details can be found in Section 3.} the  concurrency~\cite{Ruochen2020} inherent in networked discrete-event systems. 
Some  insights and contributions of this paper are shown in the following, which shall set our work apart from the existing works: 
\begin{enumerate}
     \item We model the networked closed-loop system as the synchronous product of  a transformed plant $P$ and a networked supervisor $S$.  We consider a networked supervisor as  controlling its outputs (i.e., the sending of control messages) and observing both its inputs and outputs (i.e., the receiving of observation messages and the sending of control messages). This by construction ensures the (state-based) normality property~\cite{wonham2015supervisory},~\cite{su2010model} can be applied to synthesize the unique, supremal networked supervisor. An input-output view has also been adopted in~\cite{balemi1992supervision} and~\cite{balemi1994input}, but our interpretation of inputs and outputs is fundamentally different from those of~\cite{balemi1992supervision},~\cite{balemi1994input}. It is worth mentioning that the networked supervisor considered in this paper is a command non-deterministic supervisor which has the freedom to choose among different control commands to issue at the same supervisor state~\cite{zhusupervisor}, and thus it is at least as powerful as its deterministic counterpart that are considered in the Ramadge-Wonham supervisor synthesis framework. 
     We shall not impose any assumption on the set of controllable events and the set of observable events for the original plant. We do not require each plant event to be observable, which is assumed in~\cite{rashidinejad2018supervisory} and~\cite{rashidinejad2019}.
\item   We here model the synchronous product of the original
plant, the command execution automaton and the
two channel automata as the transformed plant (see
Fig. 1). Thus, both network delays and communication losses are   embedded into the transformed plant and their adverse effects are captured within the (non-networked) observability and controllability properties~\cite{wonham2015supervisory},~\cite{su2010model} over the transformed plant. It follows that the properties of network observability and network controllability~\cite{lin2014control} and other similar extensions collapse to their non-networked counterparts over the transformed plant. In particular, by 1), observability property coincides with normality property for our setup. 
\item The solution methodology of this work is to model the networked supervisor synthesis problem as the  (non-networked) supervisor synthesis problem (for non-deterministic plants). This allows us to employ the existing non-networked supervisor synthesis algorithms to deal with the supervisor synthesis problem for networked systems when facing other kinds of specifications, such as the range control specification~\cite{Yin}, without the need to invent new synthesis algorithms (cf.~\cite{Shu2015}).  This also leads to a significant reduction in the implementation effort. In fact, we could now employ the existing (non-networked) supervisor synthesis tools to solve the networked supervisor synthesis problem for free, after a straightforward model transformation, and take advantage of the existing complexity reduction and symbolic techniques and  that have already been implemented in these tools. For example, the  supervisor reduction algorithms,  developed for the non-networked setup, could be used to synthesize the supremal networked supervisor with a reduced size~\cite{su2004supervisor} (see Section 5).
\end{enumerate}
Since the (non-networked) supervisor synthesis problem is a degenerate case of the networked supervisor synthesis problem (with zero delays and zero message losses), this paper shows that the networked supervisor synthesis problem and the non-networked supervisor synthesis problem are inter-transformable to each other. Overall, this paper could also be viewed as providing the (state-based) normality property based synthesis approach to solve the networked supervisor synthesis problem as well as the non-networked supervisor synthesis problem in a uniform manner\footnote{When there is no network delay, by construction there is no message loss; and the receiving of any message immediately follows and thus can be merged with the sending of the same message into an atomic event, recovering the non-networked setup. For more details, please see Section 3.}, in  the synthesis of the supremal command non-deterministic supervisors.

An earlier conference version of this work that only considers FIFO channels appears in~\cite{zhusupervisor}. Indeed, due to the consideration of FIFO channels in~\cite{zhusupervisor}, the networked supervisor synthesis problem studied there can be reduced to the non-networked supervisor synthesis problem over deterministic plants. The channels there
are required to be of finite capacities, which are assumed
to be (a prior) known. In addition, the command execution
automaton of this paper can avoid the following undesirable effects (of the one adopted in~\cite{zhusupervisor}) from occurring:
a) network delays can delay the execution of all the  uncontrollable events, b)  disabling the issuance of all the
control commands can disable all the uncontrollable events. This is achieved by defining a control command to only contain those controllable events that it will enable, i.e., uncontrollable events will not be included in any control command. Intuitively, control commands, which can be delayed in the reception or get lost (by the control channel) and disabled in the issuance (by the networked supervisor), can only control controllable events. On the other hand, every uncontrollable event is defined at every state of the command execution automaton.

The paper is organized as follows. First, we provide some preliminaries that are helpful for understanding this paper, in Section 2. Then, in Section 3, we model the two lossy channels with bounded network delays and provide a model of the command execution mechanism. In Section 4, we shall impose a natural restriction on the networked supervisors and transform the networked supervisor synthesis problem into the classical non-networked supervisor synthesis problem~\cite{wonham2015supervisory},~\cite{su2010model}. The case study of a mini-guideway example is then presented in Section 5 to illustrate the effectiveness of our approach. Finally, we present the conclusions and future works in Section 6. 
\section{Preliminaries}
We assume the reader to be familiar with the basic theories of  finite automata and supervisory control~\cite{wonham2015supervisory}. We shall recall some additional notations  and terminologies in the following to make the paper more self-contained.



A (non-deterministic) finite state automaton $G$ over $\Sigma$ is a 5-tuple $(Q, \Sigma, \delta, q_0, Q_m)$, where $Q$ is the finite set of states; $\delta \subseteq Q \times \Sigma \times Q$ is the transition relation; $q_0  \in Q$ is the initial state and $Q_m \subseteq Q$ is the marked state set. $G$ is determinsitic, if $\delta$ is a (partial) transition function.  When $Q_m=Q$, we shall write $G=(Q, \Sigma, \delta, q_0)$. For any two finite state automata $G_1, G_2$, we write $G_1   \lVert G_2$ to denote their synchronous product.  
  A control constraint $\mathcal{A}$ over $\Sigma$ is a tuple ($\Sigma_c,\Sigma_o$), where  $\Sigma_c\subseteq\Sigma$ denotes the subset of controllable events and $\Sigma_o\subseteq\Sigma$ denotes the subset of  observable events.
As usual, we let $\Sigma_{uo}:=\Sigma\backslash\Sigma_o$ denote the subset of unobservable events and $\Sigma_{uc}:=\Sigma\backslash\Sigma_c$ the subset of uncontrollable events. 

A  supervisor (over control constraint $\mathcal{A}$) is a 
deterministic  finite state automaton $S = (X,\Sigma,\xi,x_0,X_m)$ which satisfies both  the state controllability and the state observability properties~\cite{su2010model}. If $\Sigma_c \subseteq \Sigma_o$, the state observability property can be  replaced with the state normality property~\cite{su2010model}. The properties of state controllability, state observability and state normality generalize their language-based counterparts~\cite{wonham2015supervisory}, to the setup of controlling non-deterministic plants $G$.    
 A specification   often specifies a set of bad states to avoid in the plant $G$. For example, let $Q_{bad} \subseteq Q$ denote the set of bad states to avoid (in the plant $G$).  Several existing supervisor synthesis tools such as TCT \cite{feng2006tct}, SuSyNA \cite{su2018generalized} and Supremica~\cite{akesson2003supremica} can be used for synthesizing supervisors $S$ such that a) no state $(x, q_{bad}) \in X \times Q_{bad}$ can be reached in the closed-loop system $S \lVert G$ and b) $S \lVert G$ is non-blocking, i.e. every reachable state can reach some marked state in $S \lVert G$. Supremica is tailored for deterministic plants, while SuSyNA and TCT are able to deal with both deterministic and non-deterministic plants\footnote{TCT is mainly designed for supervisor synthesis over deterministic plants. However, it can also be adapted for  synthesis over non-deterministic plants.  Supremica does not allow non-deterministic plants as the input.}. A procedure for the synthesis of the supremal nonblocking state-normal supervisor is available in~\cite{su2010model} and has been implemented in SuSyNA.

We have the following definitions in the supervisory control and observation problem ({\bf SCOP}) of~\cite{wonham2015supervisory}, i.e., the partial-observation supervisor synthesis problem for deterministic plants in the setup that $\Sigma_c \subseteq \Sigma_o$. Let $L(G)$ denote the closed behavior of $G$ and $L_m(G)$ the marked behavior of $G$.
\begin{enumerate}
\item A language $K \subseteq \Sigma^*$ is said to be
controllable (with respect to $G$ and $\Sigma_{uc}$) if $\overline{K}\Sigma_{uc} \cap L(G)\subseteq \overline{K}$, where $\overline{K}$ denotes the prefix-closure of $K$~\cite{wonham2015supervisory}. 
The set of all  controllable sublanguages of a language $E \subseteq \Sigma^*$ is denoted by $\mathcal{C}(E)=\{K \subseteq E \mid \overline{K}\Sigma_{uc} \cap L(G) \subseteq \overline{K}\}$. 
\item $K$ is   $(L_m(G), P_0)$-normal~\cite{wonham2015supervisory} if $K=P_{0}^{-1}(P_{0}(K)) \cap L_m(G)$,  where $P_0$ and $P_0^{-1}$ denotes the natural projection and the inverse projection on languages, respectively. The set of all $(L_m(G), P_0)$-normal sublanguages of $E$ is denoted by $\mathcal{N}(E; L_m(G))=\{K \subseteq E \mid K\text{ is }(L_m(G), P_0)\text{-normal}\}$.
\item Similarly, we say $\overline{K}$ is $(L(G), P_0)$-normal~\cite{wonham2015supervisory} if $\overline{K}=P_{0}^{-1}(P_{0}(\overline{K})) \cap L(G)$. As usual, we let $\overline{\mathcal{N}}(E; L(G))=\{K \subseteq E \mid \overline{K}\text{ is }(L(G), P_0)\text{-normal}\}$.
\item Let $\mathcal{S}(E)=\mathcal{C}(E)\cap N(E; L_m(G)) \cap \overline{\mathcal{N}}(E; L(G))$~\cite{wonham2015supervisory}. $\mathcal{S}(E)$ is commonly referred to as the set of normal and controllable sublanguages of $E$~\cite{wonham2015supervisory},~\cite{ZMM05}. Since $\mathcal{S}(E)$ is non-empty (as $\varnothing \in \mathcal{S}(E)$) and closed under arbitrary unions, $\it{sup}\mathcal{S}(E)$ exists in $\mathcal{S}(E)$.
\end{enumerate}
{\bf SCOP} has  $\it{sup}\mathcal{S}(E)$ as the supremal solution, which  is known as the supremal normal and controllable sublanguage; $\it{sup}\mathcal{S}(E)$ can be synthesized via different successive approximation algorithms~\cite{wonham2015supervisory},~\cite{ZMM05},~\cite{WLLW18},~\cite{Lin2020T}. The synthesis of $\it{sup}\mathcal{S}(E)$ has already been implemented in TCT and SuSyNA; in particular, the supremal nonblocking state-normal supervisor, synthesized with SuSyNA, corresponds to $\it{sup}\mathcal{S}(E)$ when $G$ is deterministic~\cite{su2010model}.

\begin{figure}
    \centering
    \includegraphics[scale = 0.4]{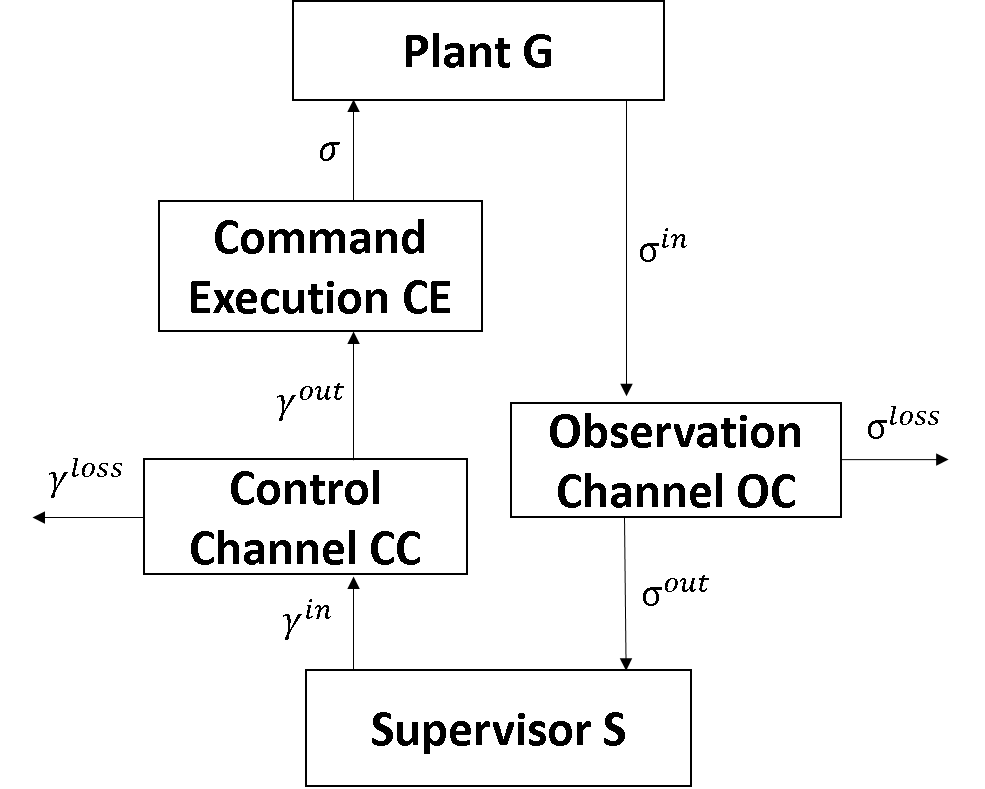}
    \caption{Networked supervisory control architecture}
    \label{fig:my_label}
\end{figure}







\section{Network Control Architecture}
In this section, we briefly talk about the network control architecture that will be followed in this work. Then, we explain how the components work and how they can be modelled. In this paper, we define the set of all possible control commands to be $\Gamma=2^{\Sigma_c}\backslash\{\varnothing\}$, deviating from the standard definition of $\Gamma$. Intuitively, in this work each control command $\gamma \in \Gamma$ only contains those controllable events that it will enable. 

The network control architecture is shown in Fig. 1. The networked closed-loop system consists of the plant $G$, the (networked) supervisor $S$, the observation channel $OC$ from the plant to the supervisor that carries the observation messages, the control channel $CC$ from the supervisor to the plant which carries the control messages, and the command execution automaton $CE$ which executes the events based on the received control messages. These five components\footnote{$G$ and $CE$ together can be viewed as an independent module; each of $OC$, $CC$ and $S$ can be viewed as an independent module.} interact with each other (through the  corresponding inputs and outputs) and form the closed-loop control system. An alternative perspective on the network control architecture which is very different from ours is available in~\cite{balemi1992supervision},  \cite{balemi1994input}. The observation channel and the control channel can be either FIFO or non-FIFO in this paper. In our approach, we only need to obtain the  automaton models of these two channels; the message ordering property of these two channels does not matter at all. For the sake of brevity, in the rest of this work, we only consider non-FIFO observation channel and FIFO control channel; both channels are assumed to be lossy.


\subsection{Observation Channel Automaton}

 Whenever the plant $G$ executes an observable event $\sigma \in \Sigma_o$, it sends the observation message $m_{\sigma}$, indicating the occurrence of $\sigma$ in the plant, over the observation channel; the event of sending the message $m_{\sigma}$ is denoted by $\sigma^{in} \in \Sigma_{o}^{in}$, where $\Sigma_{o}^{in}$ is a copy of $\Sigma_o$ with superscript $``in"$. The event $\sigma^{in}$ updates the content of the observation channel by adding message $m_{\sigma}$.  We treat\footnote{The refined semantics where $\sigma$ and $\sigma^{in}$ are treated as two distinct events, when $\sigma \in \Sigma_o$, can also be modelled.} the execution of $\sigma$ and the sending of $m_{\sigma}$, i.e., the execution of $\sigma^{in}$, together as an atomic event, if $\sigma$ is observable. We later simply identify this atomic event as $\sigma^{in}$. The event of receiving message $m_{\sigma}$ by the supervisor is denoted by $\sigma^{out} \in \Sigma_{o}^{out}$, where $\Sigma_{o}^{out}$ is a copy of $\Sigma_o$ with superscript $``out"$. The execution of $\sigma^{out}$ indicates the occurrence of $\sigma^{in} \in \Sigma_{o}^{in}$ (some moment ago), and updates the content of the observation channel by removing message $m_{\sigma}$. 
 
 Due to the non-FIFO property of the observation channel, the message  that enters the channel first may not get out first. A message will stay in the channel for some time, representing the time used for the transmission of that message, which is known as the delay. To describe the delays that may occur in the channel, the channel is assigned a fixed integer, which is referred to as the delay bound of the channel. Intuitively, any message can stay in the channel for up to the number of time steps specified by the delay bound. For each message sent over the channel, it is always coupled with the delay bound of the channel initially, which is decreased afterwards whenever some plant event is executed (i.e., the occurrence of a plant event is defined to constitute a time step). Each message must get out of the channel when this (coupled) count decreases to zero. For this non-FIFO observation channel, any message can first get out of the channel regardless of the count. 
We denote the fixed delay bound of the observation channel as $num^o$, which corresponds to the maximum number of time steps that a message can stay in the channel. Thus, each element in the channel 
is a two tuple. The first tuple is the physical message,
and the second tuple is used to record for at most how many
time steps can the message still stay in the channel. We
term the second tuple as time-to-leave for the corresponding
message. 
  The observation channel can have  infinite capacity and may experience loss of messages. We assume
messages $m_{\sigma}$, where $\sigma \in \Sigma_{ol} \subseteq \Sigma_o$, may get lost in the
observation channel. Thus,  $\Sigma_{ol}$ denotes the set of lossy (observable) events. 

The observation channel from the plant to the supervisor can be represented by an automaton $OC$ over alphabet $\Sigma_{obs}= \Sigma_{uo} \cup \Sigma_{o}^{in} \cup \Sigma_{o}^{out} \cup \Sigma_{ol}^{loss}$, where $\Sigma_{ol}^{loss}$ denotes the copy of $\Sigma_{ol}$ with superscript $``loss"$.  


The state space of the observation channel automaton is denoted as $Q^{obs}$. For each $\sigma \in \Sigma_o$,  each $0 \leq i \leq num^o$ and each $k >0$, let $mls = ((m_{\sigma}, i), k)$ denote the multi-set which consists of $k$ copies of $(m_{\sigma}, i)$, where $i$ is the time-to-leave tuple discussed before. Then, the content of the observation channel  can be represented by $mlss \subseteq \{((m_{\sigma}, i), k) \mid \sigma \in \Sigma_o, 0 \leq i \leq num^o, k>0\}$ with\footnote{In the definition for subset inclusion between collections of multi-sets, we shall treat each multi-set as an element. Thus,  $\{((m_{\sigma}, 1), 3)\} \subseteq \{((m_{\sigma}, 1), 3), ((m_{\sigma}, 3), 2)\}$, while in our convention we write $\{((m_{\sigma}, 1), 1)\} \not \subseteq \{((m_{\sigma}, 1), 3), ((m_{\sigma}, 3), 2)\}$.} each multi-set in $mlss$ having distinct $(m_{\sigma}, i)$. The set of all possible contents of the observation channel is denoted by $S_{obs}$. Then, we have 
$Q^{obs}=\{q^{mlss} \mid mlss \in S_{obs}\}$. Let $m(mlss)$ denote the minimum of the time-to-leaves for the multi-sets in $mlss$.
The  transition relation $\delta^{obs}\subseteq  Q^{obs} \times \Sigma_{obs} \times Q^{obs}$ of $OC$ is defined as follows.

\begin{enumerate}
\item For each $mlss\in S_{obs}$ and each $\sigma \in \Sigma_o$, we define
\begin{enumerate}
\item $(q^{mlss} , \sigma^{in}, q^{mlss'}) \in \delta^{obs}$, where $mlss'\in S_{obs}$ is obtained from $mlss\in S_{obs}$ by: i) replacing each $((m_{\sigma'}, i), k)$ in $mlss$ with $((m_{\sigma'}, i-1), k)$, where $\sigma' \in \Sigma_o$, and ii) adding $((m_{\sigma}, num^o), 1)$, 
if $m(mlss) \geq 1$ or  $mlss=\{\}$. 
\item $(q^{mlss} , \sigma^{out}, q^{mlss'}) \in \delta^{obs}$, where $mlss' \in S_{obs}$ is obtained from $mlss \in S_{obs}$ by replacing some  $((m_{\sigma}, i), k)$ with $((m_{\sigma}, i), k-1)$, if $k> 1$, or by removing some  $((m_{\sigma}, i), 1)$.
\item $(q^{mlss} , \sigma^{loss}, q^{mlss'}) \in \delta^{obs}$, where $\sigma \in \Sigma_{ol}$ and $mlss'$ is obtained from $mlss$ by replacing some $((m_{\sigma}, i), k)$ in $mlss$ with $((m_{\sigma}, i), k-1)$, if $k >1$, or by removing some $((m_{\sigma}, i), 1)$ in $mlss$. 
\end{enumerate}
\item For  each $mlss\in S_{obs}$ and each $\sigma \in \Sigma_{uo}$, we define
 $(q^{mlss} , \sigma, q^{mlss'}) \in \delta^{obs}$, where $mlss'\in S_{obs}$ is obtained from $mlss$ by replacing each $((m_{\sigma'}, i), k)$ in $mlss$  with $((m_{\sigma'}, i-1), k)$,  where $\sigma' \in \Sigma_o$, if $m(mlss) \geq 1$ or\footnote{If $mlss=\{\}$, then clearly $mlss'=\{\}$.} $mlss=\{\}$. 
\end{enumerate} Rule
1. a) states that event $\sigma^{in}$ constitutes a time step and  adds a message $m_{\sigma}$ coupled with the time-to-leave $num^o$. Rule 1. b) states that event $\sigma^{out}$ removes some message $m_{\sigma}$ from the observation channel. Rule 1. c) says that message $m_{\sigma}$ can get lost, when $\sigma \in \Sigma_{ol}$. Rule 2) says that any unobservable event $\sigma \in \Sigma_{uo}$  also constitutes a time step. 

Then, the observation channel automaton is given by the 4-tuple $OC=(Q^{obs}, \Sigma_{obs}, \delta^{obs}, q^{\{\}})$. In general, $OC$ is  non-deterministic  due to Rules 1. b) and 1. c) and $Q^{obs}$ has an infinite number of states. We shall now show that the set of reachable states of $OC$ is of  finite cardinality.  \\
\begin{lemma}
\label{lemma: k}
For any reachable state $q^{mlss} \in Q^{obs}$ of $OC$, it holds that, for each $((m_{\sigma}, i), k) \in mlss$,  $k=1$.
\end{lemma}
{\em Proof}: For the initial state, the stated property clearly holds. The proof for the general case straightforwardly follows by an induction on the number of transitions to reach  $q^{mlss}$. Intuitively, the stated property on a state of $Q^{obs}$ is invariant under any one-step transition of $\delta^{obs}$ according to Rules 1) and 2). \hfill \qed \\
\begin{proposition}
The reachable state set of $OC$ is of finite cardinality.
\end{proposition}
{\em Proof}: By Lemma~\ref{lemma: k}, for each reachable state $q^{mlss}$ of $OC$, we have $mlss\subseteq \{((m_{\sigma}, i), 1) \mid \sigma \in \Sigma_o, 0 \leq i \leq num^o\}$. Thus, there are only a finite number of choices of $mlss$ and it follows that the reachability set of $OC$ is of finite cardinality.  Indeed, any two  observation messages in the observation channel must have different time-to-leaves, which could be shown with a straightforward induction proof. Thus, there can be at most $num^o+1$ messages in the observation channel at any moment. The  number of reachable states for $OC$ is thus $(1+|\Sigma_o|)^{num^o+1}$.\hfill \qed \\

We remark that $num^o+1$ is indeed the tight upper bound for the channel capacity of the observation channel.

\subsection{Control Channel Automaton} 
The event that the supervisor sends a control message $m_{\gamma}$ over the control channel is denoted by $\gamma^{in} \in \Gamma^{in}$, where $\Gamma^{in}$ is a copy of $\Gamma$ with superscript $``in"$. The event $\gamma^{in}$ updates the content of the control channel by appending the message $m_{\gamma}$ to the tail of the (message) queue. The event of receiving message $m_{\gamma}$ by the plant\footnote{We shall treat $G$ and $CE$ together as the augmented plant. Whenever there is no risk of confusion, we shall also refer to $G$ and $CE$ together as the plant.} is denoted by ${\gamma}^{out} \in \Gamma^{out}$, where $\Gamma^{out}$ is a copy of $\Gamma$ with superscript $``out"$. The event  ${\gamma}^{out}$ updates the content of the control channel by popping the head of the queue. Similarly, we denote the fixed delay bound of the control channel as $num^c$. The control channel also can have infinite capacity and may also experience loss of messages. We assume each message $m_{\gamma}$, where $\gamma \in \Gamma$, may get lost in the control channel. 


The control channel from the supervisor to the plant can be represented by the automaton $CC$ over alphabet $\Sigma_{com} = \Sigma_{uo} \cup \Sigma_o^{in} \cup \Gamma^{in} \cup \Gamma^{out} \cup \Gamma^{loss}$, where $\Gamma^{loss}$ denotes the copy of $\Gamma$ with superscript $``loss"$. 

The state space of the control channel automaton is denoted as $Q^{com}$. Let $S_{com}$ denote the collection of strings over the (auxiliary) alphabet $\Sigma_{aux}=\{m_{\gamma} \mid \gamma \in \Gamma\} \times [0, num^c]$. Then, $Q^{com}=\{q^s \mid s \in S_{com}\}$. Let $m(s)$ denote the minimum of the time-to-leaves for the elements in $s$. The transition relation $\delta^{com} \subseteq Q^{com} \times \Sigma_{com} \times Q^{com}$ of $CC$ is defined as follows.

\begin{enumerate}
\item For each $s\in S_{com}$ and each $\gamma \in \Gamma$, we define
\begin{enumerate}
\item $(q^s , \gamma^{in}, q^{s'}) \in \delta^{com}$, where $s'=s(m_{\gamma}, num^c)$ is the concatenation of $s$ and $(m_{\gamma}, num^c)$
\item $(q^s , \gamma^{out}, q^{s'}) \in \delta^{com}$, where $s=(m_{\gamma}, i)s'$ for some $i \geq 0$
\item $(q^s , \gamma^{loss}, q^{s'}) \in \delta^{com}$, where $s=s_1(m_{\gamma}, i)s_2$ for some $s_1, s_2$ over $\Sigma_{aux}$  and $s'=s_1s_2$
\end{enumerate}
\item For each $s\in S_{com}$ and each $\sigma \in \Sigma_{uo}$, we define $(q^s , \sigma, q^{s'}) \in \delta^{com}$, where $s=(m_{\gamma_1}, i_1)\ldots (m_{\gamma_n}, i_n)$, for some $\gamma_1, \ldots, \gamma_n$, $i_1, \ldots, i_n$, and $s'=(m_{\gamma_1}, i_1-1)\ldots (m_{\gamma_n}, i_n-1)$, if $m(s)\geq 1$ or\footnote{If $s=\epsilon$, then clearly $s'=\epsilon$.} $s=\epsilon$. 
\item For each $s\in S_{com}$ and each $\sigma \in \Sigma_{o}$,  we define $(q^s , \sigma^{in}, q^{s'}) \in \delta^{com}$, where $s=(m_{\gamma_1}, i_1)\ldots (m_{\gamma_n}, i_n)$, for some $\gamma_1, \ldots, \gamma_n$, $i_1, \ldots, i_n$, and $s'=(m_{\gamma_1}, i_1-1)\ldots (m_{\gamma_n}, i_n-1)$, if $m(s)\geq 1$ or $s=\epsilon$. 
\end{enumerate}
Intuitively, Rule 1. a) says that $\gamma^{in}$ appends the queue with a message $m_{\gamma}$ that is coupled with the time-to-leave $num^c$. Rule 1. b) states that $\gamma^{out}$ removes the message $m_{\gamma}$ from the head of the queue. Rule 1. c) says that any message $m_{\gamma}$ can get lost. Rule 2) and Rule 3) specify the effect of the execution of a plant event to the control channel, where the physical messages are unchanged but the time-to-leaves decrease by 1. 

 The control   channel automaton is given by the 4-tuple $CC = (Q^{com}, \Sigma_{com}, \delta^{com}, q^{\epsilon})$. In general, $CC$ is  non-deterministic due to Rule 1. c) and $Q^{com}$ has an infinite number of states. 
 We  have the following.\\
 
 \begin{proposition}
 \label{proposition: bip}
 If $num^c=0$, then the set of reachable states of $CC$ is of cardinality $2^{|\Sigma_c|}$; if $num^c \geq 1$, then the set of reachable states of $CC$ is infinite.
 \end{proposition}
 {\em Proof}: Suppose $num^c=0$. Initially, the content of the control channel is empty and thus $s=\epsilon$. Only Rule 1. a), Rule 2) and Rule 3) can be applied to state $q^{\epsilon}$. Rule 2) and Rule 3) do not generate a new reachable state. Rule 1. a) generates  $2^{|\Sigma_c|}-1$ new reachable states from $q^{\epsilon}$ within one transition;  these new reachable states are in the set $\{q^{(m_{\gamma}, 0)} \mid \gamma \in \Gamma\}$. Only Rule 1. b) can be applied to the states in $\{q^{(m_{\gamma}, 0)} \mid \gamma \in \Gamma\}$, and  $q^{\epsilon}$ is reached. Thus, there are exactly $2^{|\Sigma_c|}$ reachable states. In particular, by construction, the control channel is lossless in this case.
 
 If $num^c \geq 1$, Rule 1. a) can already be applied to state $q^{\epsilon}$ for any number of times, each time generating a new reachable state. Thus, in this case, the set of reachable states of $CC$ is infinite.\qed \\
 
\subsection{Command Execution Automaton}
While the models of the observation channel $OC$ and the control channel $CC$ provide details on how the  sending, receiving and loss of messages will influence the content of the channels, there is no specification on how the supervisor and the command execution automaton will interact with the channels. In this subsection, we explain the command execution mechanism that will be considered in this paper and show how it can be modelled as an automaton. A control command $\gamma$ and the corresponding control message $m_{\gamma}$ are not the same\footnote{A control message $m_{\gamma}$ encodes the control command $\gamma$.}. For the sake of convenience, we shall use both terms interchangeably, as there is no risk of confusion.

The command execution automaton could receive multiple control commands within one time step. There could be several different ways for a command execution automaton to deal with these control commands which are received within the same time step, before any plant event is executed. For example, the command execution automaton can a) store these control commands and execute them in sequence, or b) execute the first control command that is received and throw away the rest that are received within the same time step, or c) keep updating the received control command and execute the last control command that is received  within the same time step, before a plant event is executed. In this work, we assume the second option is adopted, but we shall also explain how the third option can be easily modelled\footnote{The adoption of the second option is natural in some sense, when control messages could get lost. In particular, the supervisor may send a sequence of identical control messages within each time step to have necessary redundancy against the losses of control messages; in this case, the command execution automaton only needs to use the first received control command and throw away the rest that are received within the same time step. The adoption of the first option will cause memory blowup for the modelling of the command execution automaton and thus is not considered in this work.}. We assume each controllable event is permitted to occur only when the control command issued by the supervisor arrives (after some bounded communication delays), as in~\cite{park2002robust}, \cite{park2006delay}, \cite{park2007supervisory}, while the execution of an unobservable event in the plant will lead to the reuse of the most recently used control command; in particular, all the control messages received within the next time step will be thrown away, if any, after the execution of an unobservable event in the plant\footnote{To cater to different command execution mechanisms,  we only need to adapt the command execution automaton, without changing the  supervisor synthesis algorithm.}. 


We now create the command execution automaton $CE$, which shows the following: a) how a control command is executed, b) how to deal with the multiple control commands that are received within the same time step, and c) how to deal with the execution of an unobservable event. The command execution automaton $CE$ is given by the tuple $(Q^{CE}, \Sigma_{CE}, \delta^{CE},q_0^{CE})$, where $Q^{CE}=\{q^{\gamma} \mid \gamma \in \Gamma\} \cup \{q_{wait}\}$, $\Sigma_{CE}=\Gamma^{out} \cup \Sigma_{o}^{in} \cup \Sigma_{uo}$, $q_0^{CE}=q_{wait}$. $\delta^{CE}: Q^{CE} \times \Sigma_{CE} \rightarrow Q^{CE}$ is defined as follows. 
 \begin{enumerate}
     \item for any $\sigma \in \Sigma_{uc} \cap \Sigma_{uo}$, $\delta^{CE}(q_{wait}, \sigma)=q_{wait}$,
     \item for any $\sigma \in \Sigma_{uc} \cap \Sigma_{o}$, $\delta^{CE}(q_{wait}, \sigma^{in})=q_{wait}$,
      \item for any $\gamma \in \Gamma$, $\delta^{CE}(q_{wait},\gamma^{out})=q^{\gamma}$,
     \item for any $\gamma, \gamma' \in \Gamma$, $\delta^{CE}(q^{\gamma}, \gamma'^{out})=q^{\gamma}$,
     \item for any $q^\gamma$, if $\sigma \in \Sigma_{o} \cap (\gamma \cup \Sigma_{uc})$, $\delta^{CE}(q^{\gamma},\sigma^{in})=q_{wait}$,
      \item for any $q^\gamma$, if $\sigma \in \Sigma_{uo} \cap (\gamma \cup \Sigma_{uc})$, $\delta^{CE}(q^{\gamma},\sigma)=q^{\gamma}$,
     \item and no other transitions are defined
 \end{enumerate}

We shall now explain how $CE$ works. Intuitively, at the initial state $q_{wait}$, $CE$ waits
to receive a control message, while in the mean time any uncontrollable event can be executed and will only lead to a self-loop at $q_{wait}$. This is reflected in Rules 1), 2). Rule 3) says that once a control message $m_{\gamma}$ is received, it transits to state $q^{\gamma}$ that records this most recently received control command, which will be used next. Any other control commands received within the same time step will be ignored, leading to a self-loop at state $q^{\gamma}$, which is reflected in Rule 4). Then, only those events in $\gamma \cup \Sigma_{uc}$ are allowed to be fired. If an observable event $\sigma \in \Sigma_o \cap (\gamma \cup \Sigma_{uc})$ is fired at state $q^{\gamma}$, then $CE$ returns to the initial state $q_{wait}$, waiting to receive a new control command; if an unobservable event $\sigma \in \Sigma_{uo} \cap (\gamma \cup \Sigma_{uc})$ is fired at $q^{\gamma}$ instead, the command execution automaton self-loops at state $q^{\gamma}$ as the same control command $\gamma$ is to be used for the next event execution. This is reflected in Rule 5) and Rule 6), respectively. In particular, if an unobservable event $\sigma$ is fired at state $q^{\gamma}$, the control commands received within the next time step will be thrown away, as reflected in Rule 4) and Rule 6) combined. The state size of $CE$ is $2^{|\Sigma_c|}$. 

We here remark that, if the third option, i.e., option c), of the above-mentioned command execution mechanism is adopted, then only Rule 4) needs to be changed to 
\begin{enumerate}
    \item [4')]  for any $\gamma, \gamma' \in \Gamma$, $\delta^{CE}(q^{\gamma}, \gamma'^{out})=q^{\gamma'}$.
\end{enumerate}
Intuitively, only the most recently received control command will be stored, before a plant event is executed. 


\subsection{Relabelled Plant}
As we have discussed before, we treat both $\sigma$ and $\sigma^{in}$ together as an atomic event, when $\sigma \in \Sigma_{o}$; the resulting atomic event is denoted as $\sigma^{in}$. We need to perform a relabelling on the plant $G=(Q, \Sigma, \delta, q_0, Q_m)$ to reflect this modelling choice.  We denote the relabelled plant as $G^{mod} = (Q^{mod}, \Sigma_{mod}, \delta^{mod}, q_0^{mod}, Q_m^{mod})$, where  $Q^{mod}=Q$,  $\Sigma_{mod}=\Sigma_{uo}\cup \Sigma_{o}^{in}$, $q_o^{mod}=q_0$ and $Q_m^{mod}=Q_m$. For any $\sigma \in \Sigma_{uo}$, $\delta^{mod}(q,\sigma)=q'$ iff  $\delta(q,\sigma)=q'$. For any $\sigma \in \Sigma_o$, $\delta^{mod}(q,\sigma^{in})=q'$ iff  $\delta(q,\sigma)=q'$. Intuitively, if the label of a transition is an observable event $\sigma \in \Sigma_o$ in the original plant $G$, the transition will be relabelled by $\sigma^{in}$ in $G^{mod}$. In the rest of this work, we renew $G:=G^{mod}$, in accordance with Fig.~\ref{fig:my_label}. Thus, $G$ is over $\Sigma_{uo} \cup \Sigma_{o}^{in}$.

\section{Networked Supervisor Synthesis}
\label{section: SS}
In this section, we specify the behavior of the last component, i.e., the (networked) supervisor, impose a natural restriction and then provide the reduction-based approach for networked supervisor synthesis.

\subsection{Synthesis Algorithm}

The networked supervisor receives observation messages from the observation channel and sends control messages to the control channel. Thus, the networked supervisor $S$ is over the control constraint $(\Gamma^{in}, \Gamma^{in} \cup \Sigma_o^{out})$, i.e., $S$ controls $\Gamma^{in}$ and observes\footnote{From the networked supervisor’s point of view, only those events involving the interaction with the networked supervisor are treated as observable.} both $\Gamma^{in}$ and $\Sigma_o^{out}$. The closed-loop system in the networked setup is the synchronous product $OC\lVert CC \lVert CE \lVert G \lVert S$ (cf. Fig.~\ref{fig:my_label}). We now can view $P=OC\lVert CC \lVert CE \lVert G$ as the transformed plant over the alphabet $\Sigma_{uo} \cup \Sigma_o^{in} \cup \Sigma_o^{out} \cup \Sigma_{ol}^{loss}\cup \Gamma^{in} \cup \Gamma^{out} \cup \Gamma^{loss}$, which is then  controlled by $S$ over $(\Gamma^{in}, \Gamma^{in} \cup \Sigma_o^{out})$. Suppose we would like to enforce the state avoidance property (in addition to the non-blockingness property). Let $Q_{bad} \subseteq Q$ denote the set of bad states to avoid in the plant $G$. Then, the transformed specification  will specify the avoidance of  $Q_{bad}$ in $G$ states, in the modular presentation of $P$.  Since  $\Gamma^{in} \subseteq \Gamma^{in} \cup \Sigma_o^{out}$, the supremal solution  exists and can be computed using the supremal nonblocking state-normal supervisor synthesis procedure of~\cite{su2010model}, which has been implemented as the {\bf make\_supervisor} operation in SuSyNA. If $P$ turns out to be deterministic, then the {\bf supscop} operator of TCT that computes the supremal normal and controllable sublanguage
$\it{sup}\mathcal{S}(E)$ can also be used, where $E$ specifies the avoidance of $Q_{bad}$ in the $G$ states. It is important to remark that, here we implicitly consider command non-deterministic supervisors that can choose different control commands to send at the same supervisor state. Since the supervisor can only observe events in $\Gamma^{in} \cup \Sigma_o^{out}$, the non-determinism due to the non-FIFO control channel can be subsumed by the partial-observation. That is, if we consider FIFO observation channel and non-FIFO control channel, TCT can be used for the synthesis, without any adapation.

If $num^c=0$, i.e., when the control channel involves no delay and is thus lossless, then the above approach works since the reachable state set of $P$ is finite. Indeed, the size of the reachable state set of $P$ is upper bounded by $|Q|4^{|\Sigma_c|}(1+|\Sigma_o|)^{num^o+1}$. However, if $num^c \geq 1$, then the reachable state set of $P$ is infinite, which renders the above approach ineffective. For the rest of this section, we shall consider a natural restriction to ensure an effective procedure for the case when $num^c \geq 1$. 


First, we restrict our attention to those  supervisors that send at most $k$ control messages after receiving each $m_{\sigma}$ from the observation channel, for some $k \geq 1$. We write $S=S^k \lVert S^f$, where $S^k$ denotes the  part of the supervisor that counts (and controls) the number of control messages that has been sent (and to be sent) to be within $k$ after receiving each observation message, and $S^f$ denotes the part of the supervisor that is over $(\Gamma^{in}, \Gamma^{in} \cup \Sigma_o^{out})$, but otherwise unconstrained. It is then $S^f$ that we need to synthesize. 

The model of $S^k$ basically serves as a counter. Let $S^k=(Q^k, \Sigma_k, \delta^k, q_0^k)$, where $Q^k=\{q_{k, 0}, q_{k, 1}, \ldots, q_{k, k}\}$, $\Sigma_k=\Gamma^{in} \cup \Sigma_o^{out}$, $q_0^{k}=q_{k, 0}$ and $\delta^k: Q^k \times \Sigma_k \longrightarrow Q^k$ is defined as follows.
\begin{enumerate}
    \item for any $i \in [0, k-1]$,  $\gamma^{in} \in \Gamma^{in}$, $\delta^k(q_{k, i}, \gamma^{in})=q_{k, i+1}$
    \item for any $i \in [0, k]$ and for any $\sigma \in \Sigma_o$, $\delta^k(q_{k, i}, \sigma^{out})=q_{k, 0}$.
\end{enumerate}
Intuitively, state $q_{k, i}$ represents that $i$ control messages has been sent in the current iteration. We notice that the state size of $S^k$ is $k+1$. We have the next useful result that guarantees the effectiveness of our approach. 

\noindent
\begin{proposition}
Let $P^k=OC \lVert CC \lVert CE \lVert G \lVert S^k$. The set of reachable states of $P^k$ is of finite cardinality.
\end{proposition}
{\em Proof}: We only need to track the length of the queue in the control channel in $P^k$ and show it is upper bounded by a constant. It then follows that the reachable state set of $P^k$ is finite. We use inductive invariants on states of $S^k$ to establish the proof. 

At any state of $P^k$, let $l_{CC}$ denote the length of the queue in the control channel and let $l_{OC}$ denote the number of observation messages in the observation channel. And, let $m_{ttl}$ denote the minimum time-to-leave of the control messages in the control channel. We claim that at each state $q_{k, i}$ of $S^k$, $l_{CC}+(l_{OC}+m_{ttl})k \leq (num^o+num^c+1)k+i$ is an inductive invariant, where $m_{ttl}$ is defined to be zero if the queue is empty. We recall that $P^k$ is over $\Sigma_{uo} \cup \Sigma_o^{in} \cup \Sigma_o^{out} \cup \Sigma_{ol}^{loss}\cup \Gamma^{in} \cup \Gamma^{out} \cup \Gamma^{loss}$.

At the initial state of $P^k$, where $S^k$ is at the initial state $q_{k,0}$, $l_{CC}=l_{OC}=m_{ttl}=0$; thus $l_{CC}+(l_{OC}+m_{ttl})k \leq (num^o+num^c+1)k$  holds. Let $p_0$ be any given state of $P^k$, where $S^k$ is at the initial state $q_{k, 0}$. Suppose $l_{CC}+(l_{OC}+m_{ttl})k \leq (num^o+num^c+1)k$  holds. We shall examine each event in $\Sigma_{uo} \cup  \Sigma_o^{in} \cup \Sigma_o^{out} \cup \Sigma_{ol}^{loss}\cup  \Gamma^{in} \cup \Gamma^{out} \cup \Gamma^{loss}$ that may be defined at $p_0$. Only those events in $\Sigma_{uo} \cup  \Sigma_o^{in} \cup \Sigma_o^{out} \cup \Sigma_{ol}^{loss} \cup \Gamma^{out} \cup \Gamma^{loss}$ will maintain $S^k$  at the initial state $q_{k,0}$, if they are defined. For each $\sigma \in \Sigma_{uo}$ that is defined at $p_0$,  $m_{ttl}$ will (either) decrease by 1  or remain the same (if the queue is empty), and thus the inductive invariant holds. For each $\sigma^{in} \in \Sigma_o^{in}$ that is defined at $p_0$, $l_{OC}$ will increase by 1 and $m_{ttl}$ will decrease by 1 or remain the same (if the queue is empty); thus, the inductive invariant holds. For each $\sigma^{out}\in \Sigma_o^{out}$ and each $\sigma^{loss} \in \Sigma_{ol}^{loss}$ that is defined at $p_0$, $l_{OC}$ will decrease by 1 and thus the inductive invariant holds. And, for each $\gamma^{out} \in \Gamma^{out}$ and each $\gamma^{loss}\in \Gamma^{loss}$ that is defined at $p_0$, $l_{CC}$ will decrease by 1 and thus the inductive invariant holds. 

For each $\gamma^{in} \in \Gamma^{in}$ that is defined at state $p_0$, state $q_{k, 1}$ is reached in $S^k$ and $l_{CC}$ will increase by 1. Thus, inductive invariant $l_{CC}+(l_{OC}+m_{ttl})k \leq (num^o+num^c+1)k+1$ will now be maintained at the reached state $p_1$. Similarly, only those events in $\Sigma_{uo} \cup  \Sigma_o^{in} \cup \Sigma_o^{out} \cup \Sigma_{ol}^{loss} \cup \Gamma^{out} \cup \Gamma^{loss}$ will maintain $S^k$  at the state $q_{k,1}$, if they are defined, which will maintain the inductive invariant $l_{CC}+(l_{OC}+m_{ttl})k \leq (num^o+num^c+1)k+1$.  For each $\gamma^{in} \in \Gamma^{in}$ that is defined at state $p_1$, state $q_{k, 2}$ is reached in $S^k$ and $l_{CC}$ will increase by 1. Thus, the inductive invariant $l_{CC}+(l_{OC}+m_{ttl})k \leq (num^o+num^c+1)k+2$ will be maintained at the reached state $p_2$. Following the same reasoning, the inductive invariant $l_{CC}+(l_{OC}+m_{ttl})k \leq (num^o+num^c+1)k+k$ will be maintained at the reached state $p_k$ when $S^k$ is at state $q_{k, k}$. Lastly, only those events in $\Sigma_{uo} \cup  \Sigma_o^{in} \cup \Sigma_{ol}^{loss} \cup \Gamma^{out} \cup \Gamma^{loss}$, if they are defined, will maintain $S^k$  at the state $q_{k,k}$ and thus the inductive invariant $l_{CC}+(l_{OC}+m_{ttl})k \leq (num^o+num^c+1)k+k$ holds;  $\gamma^{in} \in \Gamma^{in}$ cannot be executed when $S^k$ is at state $q_{k, k}$ by construction. For each $\sigma^{out} \in \Sigma_o^{out}$ defined at $p_k$, $l_{OC}$ will decrease by 1 and thus the inductive invariant $l_{CC}+(l_{OC}+m_{ttl})k \leq (num^o+num^c+1)k+k-k=(num^o+num^c+1)k$ is maintained when $S^k$ again reaches state $q_{k, 0}$. This completes the proof of the  validity of the inductive invariants. 

Now, at any reachable state $p$ of $P^k$, where $S^k$ must be at some $q_{k, i}$ state, we know that $l_{CC}+(l_{OC}+m_{ttl})k \leq (num^o+num^c+1)k+k$. Correspondingly, we know that $l_{CC} \leq (num^o+num^c+1)k+k=(num^o+num^c+2)k$, which is achieved when  $l_{OC}=m_{ttl}=0$. This completes the proof that the set of reachable states of $P^k$ is of finite cardinality, by upper bounding the length of the queue in the control channel to be within $(num^o+num^c+2)k$.

Finally, we show that there exists some $G$ for which the upper bound $(num^o+num^c+2)k$ on the length of the queue in the control channel is indeed tight.  Indeed, the string
$s=s_{-1}s_0s_1\ldots s_{num^o+num^c+1}$ over $\Sigma_{uo} \cup \Sigma_o^{in} \cup \Sigma_o^{out} \cup \Sigma_{ol}^{loss}\cup \Gamma^{in} \cup \Gamma^{out} \cup \Gamma^{loss}$ can be verified to indeed result in exactly $(num^o+num^c+2)k$  messages in the control channel, where
\begin{enumerate}
\item $s_{-1}=\sigma_1^{in}\sigma_2^{in}\ldots \sigma_{num^o+1}^{in}$, where  $\sigma_i^{in} \in \Sigma_o^{in}$ for each $i \in [1, num^o+1]$, corresponds to the phase when uncontrollable and observable events $\sigma_i$ occur in $G$ without $G$ receiving control messages
\item $s_0=\gamma_1^{in}\gamma_2^{in}\ldots \gamma_k^{in}$, where $\gamma_i^{in}\in \Gamma^{in}$ for each $i \in [1, k]$, corresponds to the phase when the supervisor issues $k$ control messages $\gamma_1^{in}\gamma_2^{in}\ldots \gamma_k^{in}$ 
\item $s_1=\sigma_1^{out}\gamma_{1,1}^{in}\ldots \gamma_{1,k}^{in}$, where $\sigma_1^{out}$ corresponds to $\sigma_1^{in}$ of $s_0$ and $\gamma_{1, i}^{in} \in \Gamma^{in}$ for each $i \in [1, k]$,  corresponds to the phase when the observation  message $m_{\sigma_1}$ is received by the supervisor, and the supervisor then issues $k$ control messages $\gamma_{1,1}^{in}\ldots \gamma_{1,k}^{in}$; similarly,  

\item  $s_2=\sigma_2^{out}\gamma_{2, 1}^{in}\ldots \gamma_{2, k}^{in}$

\item $\dots$

\item  $s_{num^o+1}=\sigma_{num^o+1}^{out}\gamma_{num^o+1,1}^{in}\ldots\gamma_{num^o+1,k}^{in}$;  after the execution of $s_{num^o+1}$,  all the $m_{\gamma_{i, j}}$'s, where $i \in [1, num^o+1], j \in [1, k]$, and all the $m_{\gamma_i}$'s, where $i \in [1, k]$, have time-to-leave $num^c$
\item $s_{num^o+2}=\sigma_{num^o+2}^{in}\sigma_{num^o+2}^{out}\gamma_{num^o+2, 1}^{in}\ldots \gamma_{num^o+2, k}^{in}$, where $\sigma_{num^o+2}^{in} \in \Sigma_o^{in}$ and $\sigma_{num^o+2}^{out} \in \Sigma_o^{out}$, and $\gamma_{num^o+2, i}^{in} \in \Gamma^{in}$ for each $i \in [1, k]$, corresponds to the phase when an uncontrollable and observable event $\sigma_{num^o+2}$ is executed in $G$ with the sent observation message $m_{\sigma_{num^o+2}}$, and next the observation message $m_{\sigma_{num^o+2}}$ is received (by the supervisor) and then the supervisor issues $k$ control messages $\gamma_{num^o+2, 1}^{in}\ldots \gamma_{num^o+2, k}^{in}$; we remark that, after the execution of  $s_{num^o+2}$, the minimum time-to-leave (of the control messages) in the control channel is $num^c-1$;  similarly,
\item $s_{num^o+3}=\sigma_{num^o+3}^{in}\sigma_{num^o+3}^{out}\gamma_{num^o+3, 1}^{in}\ldots \gamma_{num^o+3, k}^{in}$; after the execution of  $s_{num^o+3}$, the minimum time-to-leave in the control channel is  $num^c-2$
\item $\ldots$
\item \noindent $s_{num^o+num^c+1}=$ \begin{center}$\sigma_{num^o+num^c+1}^{in}\sigma_{num^o+num^c+1}^{out}\gamma_{num^o+num^c+1, 1}^{in}\ldots \gamma_{num^o+num^c+1, k}^{in}$; \end{center}
after the execution of  $s_{num^o+num^c+1}$, the minimum time-to-leave in the control channel is  $0$
\end{enumerate}

Thus, we can see that the upper bound is achieved when $G$ has an uncontrollable path defined  with length $num^o+num^c+1$ (starting from the initial state). This completes the proof.
\qed

Thus, one can effectively construct a finite state automaton model of the control channel $CC$, by assuming the channel capacity to be $(num^o+num^c+2)k$. The resultant channel model $CC^{(num^o+num^c+2)k}$, which can be effectively constructed, can be used to replace $CC$ in the construction of $P^k$. The proof of Proposition 3 ensures the soundness of the substitution. In the rest of this work, we still write $CC$ to refer to the control channel, but we will use $CC^{(num^o+num^c+2)k}$ implicitly for computation. The state size of  $CC^{(num^o+num^c+2)k}$ is (no more than) 
\begin{center}
$\sum_{j=0}^{(num^o+num^c+2)k}(2^{|\Sigma_c|}-1)^j
\binom{num^c+j}{j}\leq (2^{|\Sigma_c|}-1)^{(num^o+num^c+2)k}\sum_{j=0}^{(num^o+num^c+2)k}\binom{num^c+j}{j}=(2^{|\Sigma_c|}-1)^{(num^o+num^c+2)k}\binom{num^c+(num^o+num^c+2)k+1}{(num^o+num^c+2)k}$.
\end{center}
There are still two troubles, which are explained as follows, that often lead to an empty (supremal) networked supervisor being synthesized. 
\begin{enumerate}
\item 
It is possible that 
no control message is received by the plant due to the uncontrollable losses of control messages.
\item It is possible that 
no observation message is received by the networked  supervisor due to the uncontrollable losses of observation messages.  
\end{enumerate}
Both of these two troubles could force the networked supervisor to be pruned aggressively by the non-networked supervisor synthesis algorithm~\cite{su2010model}, in order to avoid the undesirable deadlocks, i.e., blocking, caused by the uncontrollable losses of all the control messages and observation messages,  resulting in an empty (supremal) networked supervisor to be synthesized, i.e., no solution. In the rest of this paper, we shall impose two assumptions to avoid the effect of aggressive pruning due to the uncontrollable losses of all the control messages and observation messages.

The first assumption we impose is the ($m$-)bounded consecutive losses of control messages~\cite{lin2014control}, where the control channel can have at most $m$ consecutive losses of control  messages, for some $m\geq 0$.  Intuitively, this assumption ensures the control channel eventually progresses under the losses of control messages.

The second assumption is what we shall refer to as the {\it eventual observability} property, which is formalized below.\\
\begin{definition}
Let $G=(Q, \Sigma, \delta,q_0, Q_m)$. $G$ is said to be eventually observable with respect to $\Sigma_o, \Sigma_{ol}$ if, for any $q, q' \in Q$, $\sigma \in \Sigma_{ol}$ with $\delta(q, \sigma)=q'$, it holds that 
\begin{enumerate}
    \item there exists some $ s\in  \Sigma_{uc}^*$, such that $\delta(q', s)! $ and $ s=s_1\sigma's_2$ for some $s_1, s_2 \in \Sigma_{uc}^*$, $\sigma' \in \Sigma_{uc} \cap (\Sigma_o\backslash \Sigma_{ol})$  \item for any $s \in \Sigma_{uc}^*$,  such that $\delta(q', s)!$ and $\neg \delta(q',s\sigma'')!$ for any $\sigma''\in \Sigma_{uc}$, it holds that $s=s_1\sigma's_2 $ for some $s_1, s_2\in \Sigma_{uc}^*, \sigma' \in \Sigma_{uc}\cap (\Sigma_o\backslash \Sigma_{ol})$.
\end{enumerate}
\end{definition}
Intuitively, the above   eventual observability property is imposed in this paper to guarantee the observation channel eventually progresses  under the losses of observation messages. It states that any loss of observation message can always be eventually compensated with another observation message that cannot get lost, even if no control message is received in between. 

With the above discussions, 
we now model the assumption of $m$-bounded consecutive losses of control messages with the automaton $A^m$. Here, the model of $A^m$  serves as a counter. Let $A^m=(Q^m, \Sigma_m, \delta^m, q_0^m)$, where $Q^m=\{q_{m, 0}, q_{m, 1}, \ldots, q_{m, m}\}$, $\Sigma_m=\Gamma^{loss} \cup \Gamma^{out}$, $q_0^{m}=q_{m, 0}$ and $\delta^m: Q^m \times \Sigma_m \longrightarrow Q^m$ is defined as follows.
\begin{enumerate}
    \item for any $i \in [0, m-1]$,  $\gamma^{loss} \in \Gamma^{loss}$, $\delta^m(q_{m, i}, \gamma^{loss})=q_{m, i+1}$
    \item for any $i \in [0, m]$, $\gamma^{out} \in \Gamma^{out}$, $\delta^m(q_{m, i}, \sigma^{out})=q_{m, 0}$.
\end{enumerate}
Intuitively, state $q_{m, i}$ represents that $i$ control messages have been consecutively lost in the current round. The state size of $A^m$ is $m+1$. Unfortunately, if $k \leq m$, then it is possible for all $k$ control messages to get lost. This  can be avoided by requiring $k \geq m+1$. Thus, without loss of generality, in this work we shall assume  $k=m+1$. One immediate observation is that if the supervisor sends the same control message $m+1$ times within each iteration, the control channel becomes effectively lossless. However, we still allow the supervisor to send $m+1$  generally different control messages, as long as they are permitted by the non-networked supervisor synthesis algorithm. We  remark that the eventual observability property is imposed on $G$, which does not require us to model a new component.

The networked closed-loop system is then $$OC \lVert CC \lVert A^m \lVert CE \lVert G \lVert S^{m+1} \lVert S^f.$$  We can now view $P^{m+1} \lVert A^m=OC \lVert CC \lVert A^m \lVert CE \lVert G \lVert S^{m+1}$ as the transformed plant over the alphabet $\Sigma_{uo} \cup \Sigma_o^{in} \cup \Sigma_o^{out} \cup \Sigma_{ol}^{loss}\cup \Gamma^{in} \cup \Gamma^{out} \cup \Gamma^{loss}$, which is then controlled by $S^f$ over the control constraint $(\Gamma^{in}, \Gamma^{in} \cup \Sigma_o^{out})$.
By Proposition 3, the set of reachable states of $P^{m+1} \lVert A^m$ is of finite cardinality; so, the {\bf make\_supervisor} operation of SuSyNA can be used to synthesize $S^f$. The size of the reachable state set of $P^{m+1} \lVert A^m$ is indeed upper bounded by
\begin{center}
$|Q|(m+1)(m+2)2^{|\Sigma_c|}(1+|\Sigma_o|)^{num^o+1}(2^{|\Sigma_c|}-1)^{(num^o+num^c+2)(m+1)}\binom{num^c+(num^o+num^c+2)(m+1)+1}{(num^o+num^c+2)(m+1)}$
\end{center}
It is possible to upper bound the binomial coefficient using the inequality $\binom{n}{k}\leq (\frac{en}{k})^k$ and obtain the simplified  upper bound
\begin{center}
$|Q|(m+1)(m+2)2^{|\Sigma_c|}(1+|\Sigma_o|)^{num^o+1}((2^{|\Sigma_c|}-1)e(1+\frac{1}{m+1}))^{(num^o+num^c+2)(m+1)}$.
\end{center}
\noindent
\begin{remark}
At the initial state $q_{m+1, 0}$ of $S^{m+1}$, $S^{m+1}$ can issue $m+1$ control messages to the control channel and reach the state $q_{m+1, m+1}$. By the assumption of bounded consecutive losses of control messages for the control channel, at least one control message can be received by the command execution automaton, which must be at some $q^{\gamma}$ state. If an observable plant event can later be executed in $G$ under $\gamma$, then, by the eventual observability property, at least one observation message can be received by the networked supervisor later, that is, state $q_{m+1, 0}$ can be reached in $S^{m+1}$ again. 
\end{remark}
\noindent
\begin{remark}
We emphasize that the above two assumptions are not needed for the synthesized networked supervisor to be correct. They are imposed in this work as we follow the model of time of~\cite{lin2014control}. In particular, they can be removed once we explicitly model the progression of time using the tick event~\cite{brandin1994supervisory}, which is observable to the networked supervisor and can trigger the sending of control messages to avoid the undesirable deadlocks (caused by the uncontrollable losses of control messages and observation messages).
\end{remark}
\subsection{Correctness of Synthesis}
It is clear that, by using the  non-networked supervisor synthesis algorithm, the properties of controllability, observability, nonblockingness, safety and supremality can be achieved with respect to the transformed plant, which is the physical plant that incorporates channels into its behavior modelling. It follows that the correctness of our synthesis approach solely relies on the correctness of the transformed plant. Since the models of $G$, $OC$, $CC$ are quite straightforward, the remaining task is to ensure the correctness of the command execution automaton $CE$, which models the intended command execution mechanism. 

We now explain why the same guarantees of controllability, observability, nonblockingness, safety and supremality can be ensured on the original plant $G$, with the command execution automaton $CE$,  which is quite straightforward to reason about once we slightly change our perspective. Let us first consider the case when $num^c=0$, and $OC \lVert CC \Vert CE \lVert G \lVert S$ is the networked closed-loop system. Indeed, we can view the networked closed-loop system $OC\lVert CC \lVert CE \lVert G \lVert S$ as $G \lVert (OC\lVert CC \lVert CE \lVert S)$, where $OC\lVert CC \lVert CE \lVert S$ is now the non-deterministic  supervisor that directly controls\footnote{In this viewpoint, the supervisor is non-deterministic while the plant is deterministic. We here remark that the properties of controllability and observability require special treatment, since each plant event $\sigma$ is assumed to also consume one time step, following~\cite{lin2014control}. 
} $G$. 
\begin{enumerate}
    \item 
It holds that every uncontrollable plant event is defined at each state of the supervisor $OC\lVert CC \lVert CE \lVert S$, except when being preempted by the reception of the messages whose time-to-leave items are zero. This ensures the controllability w.r.t. the plant $G$. 
    \item 
    It holds that, any unobservable plant event, if defined at a state of $OC\lVert CC \lVert CE \lVert S$, leads to a self-loop in $CE$ and $S$ and will only update the time-to-leave items of $OC$ and $CC$, without changing the physical messages within these two channels. This ensures the observability w.r.t. $G$.
    \item 
    To see that  nonblockingness, safety and supremality are also ensured w.r.t. $G$:
    \begin{enumerate} 
    \item nonblockingness of $G \lVert (OC\lVert CC \lVert CE \lVert S)$ holds since $OC\lVert CC \lVert CE \lVert G \lVert S$ is nonblocking.
    \item the bad states in $G$ are not reachable under the control of $OC\lVert CC \lVert CE \lVert S$ by construction, so safety is ensured. \item supremality of $OC\lVert CC \lVert CE \lVert S$ is ensured since $S$ is supremal by construction. 
    \end{enumerate}
\end{enumerate}
\noindent
Similar analysis can be applied to the case when $num^c \geq 1$ and $OC \lVert CC \lVert A^m \lVert CE \lVert G \lVert S^{m+1} \lVert S^f$ is the networked closed-loop system. In particular, $$OC \lVert CC \lVert A^m \lVert CE \lVert S^{m+1} \lVert S^f$$ could be viewed as the non-deterministic supervisor over $G$.
\section{A Mini-guideway Example}
\begin{figure}
    \centering
    \includegraphics[scale = 0.4]{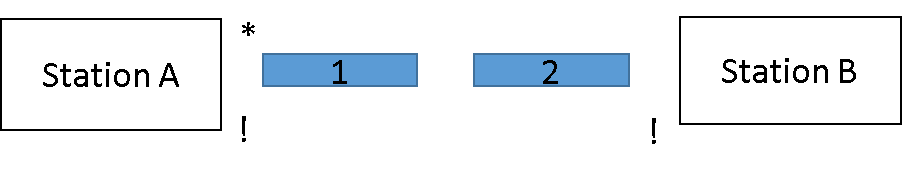}
    \caption{A mini-guideway example}
    \label{fig:guideway}
\end{figure}

In this section, we present our networked supervisor synthesis experiment on a mini-guideway example adapted from~\cite{wonham2015supervisory}. In the mini-guideway example, stations $A$ and $B$ are connected by a single one-way
track from $A$ to $B$. The track consists of 2 sections, with stoplights ($*$) and
detectors ($!$) installed at some junctions as displayed in Fig.~\ref{fig:guideway}. Two trains need to travel from $A$ to $B$, and avoid collision due to the simultaneous occupation of the same track. Due to space limitation, we shall only present the figures of  the models after the proposed transformation. We only remark here that the stoplight can decide which train can pass through a junction, while the detector can detect which train passes through the junction.

\begin{figure}
    \centering
    \includegraphics[scale = 0.4]{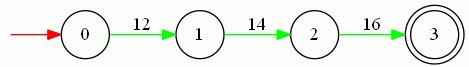}
    \caption{The relabelled automaton model $V_1^{mod}$ of train $V_1$}
    \label{fig:V1}
\end{figure}

\begin{figure}
    \centering
    \includegraphics[scale = 0.4]{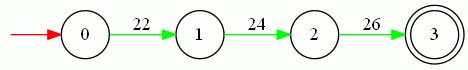}
    \caption{The relabelled automaton model $V_2^{mod}$ of train $V_2$}
    \label{fig:V2}
\end{figure}

The relabelled automaton models of the two trains are shown in Fig.~\ref{fig:V1} and Fig.~\ref{fig:V2}, respectively. Events 12, 14, 16 represent train $V_1$ crosses junctions StationA-Track1 (and the sensor sends the  observation message), Track1-Track2, Track2-StationB (and the sensor sends the  observation message), respectively;   events 22, 24, 26 represent train $V_2$ crosses junctions StationA-Track1 (and the sensor sends the  observation message), Track1-Track2, Track2-StationB (and the sensor sends the  observation message), respectively. We here remark that for the relabelled plant  $G=V_1^{mod}\lVert V_2^{mod}$, all events 12, 14, 16,  22, 24, 26 are uncontrollable and unobservable. The relabelled plant will be controlled via the control messages issued by the networked supervisor (through the command execution automaton), while any observation of event execution in the plant can only be performed indirectly through the observation messages received by the networked supervisor. This is in contrast with the unmodified model in~\cite{wonham2015supervisory}, where the events  $V_1, V_2$ crossing junction StationA-Track1 are  controllable and observable and the events $V_1, V_2$ crossing junction Track2-StationB are  observable.

\begin{figure}
    \centering
    \includegraphics[scale = 0.4]{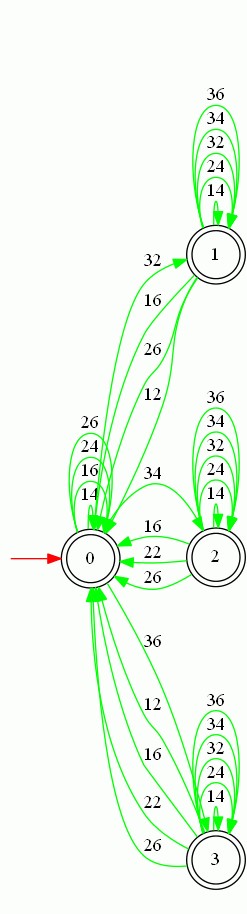}
    \caption{The command execution automaton $CE$}
    \label{fig:CE}
\end{figure}

The command execution automaton $CE$ is shown in Fig.~\ref{fig:CE}. Event 32 represents $CE$ receives a  control message that only allows train $V_1$ to cross junction StationA-Track1; event 34 represents $CE$ receives a control message that only allows train $V_2$ to cross junction StationA-Track1; event 36 represents $CE$ receives a control message that allows either $V_1$ or $V_2$ to cross the junction. 

\begin{figure*}
    \centering
    \includegraphics[width=6.8in, height= 5.0in]{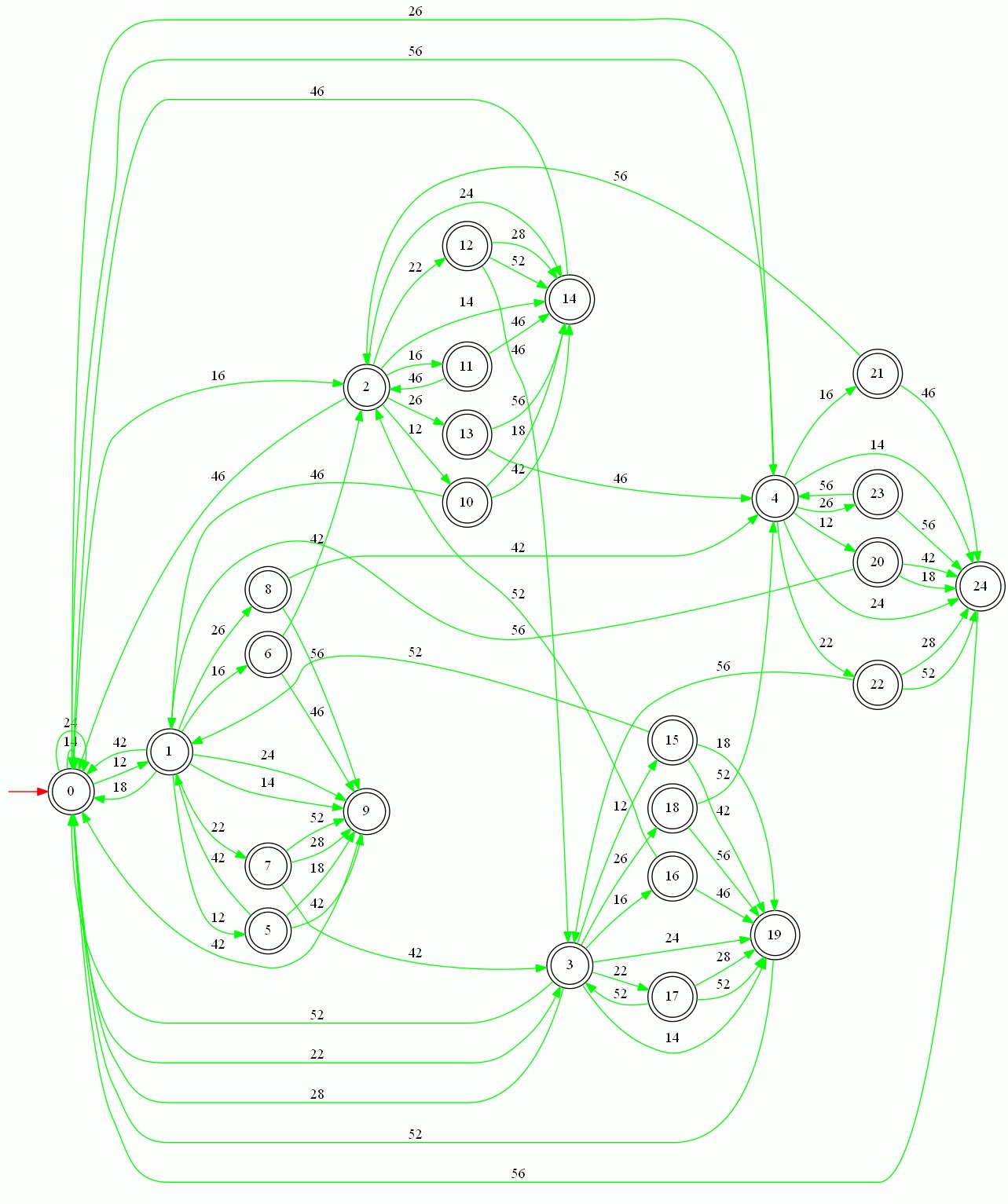}
    \caption{The observation channel automaton $OC$}
    \label{fig:OC}
\end{figure*}

\begin{figure}
    \centering
    \includegraphics[scale = 0.4]{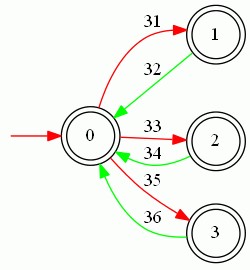}
    \caption{The control channel automaton $CC$}
    \label{fig:CC}
\end{figure}

We shall assume that the observation messages sent by the sensor at junction StationA-Track1 can get lost in the observation channel. We also assume the delay bound for the control channel is $num^c=0$, implying that the control channel is  lossless, while the delay bound for the observation channel is assumed to be $num^o=1$. The observation channel automaton $OC$ is shown in Fig.~\ref{fig:OC}, where event 18 (respectively, 28) represents the loss of the observation message that $V_1$ (respectively $V_2$) crosses junction StationA-Track1; event 42 (respectively, 52) represents the networked supervisor receives the observation message that $V_1$ (respectively, $V_2$) crosses junction StationA-Track1; event 46 (respectively, 56) represents the networked supervisor receives the observation message that $V_1$ (respectively, $V_2$) crosses junction Track2-StationB. The control channel automaton $CC$ is shown in Fig.~\ref{fig:CC}. Event 31 (respectively, 33) represents the networked supervisor sends a control message that only allows train $V_1$ (respectively, $V_2$) to cross junction StationA-Track1. Event 35 represents the networked supervisor sends a control message that allows either $V_1$ or $V_2$ to cross the junction.

Therefore, we have $\Gamma^{in}=\{31, 33, 35\}$, $\Gamma^{out}=\{32, 34, 36\}$,  $\Gamma^{loss}=\varnothing$; $\Sigma_o^{in}=\{12, 22, 16, 26\}$, $\Sigma_o^{out}=\{42, 52, 46, 56\}$,  $\Sigma_{ol}^{loss}=\{18, 28\}$ and $\Sigma_{uo}=\{14, 24\}$.

\begin{figure*}

 \center

  \includegraphics[width=\textwidth]{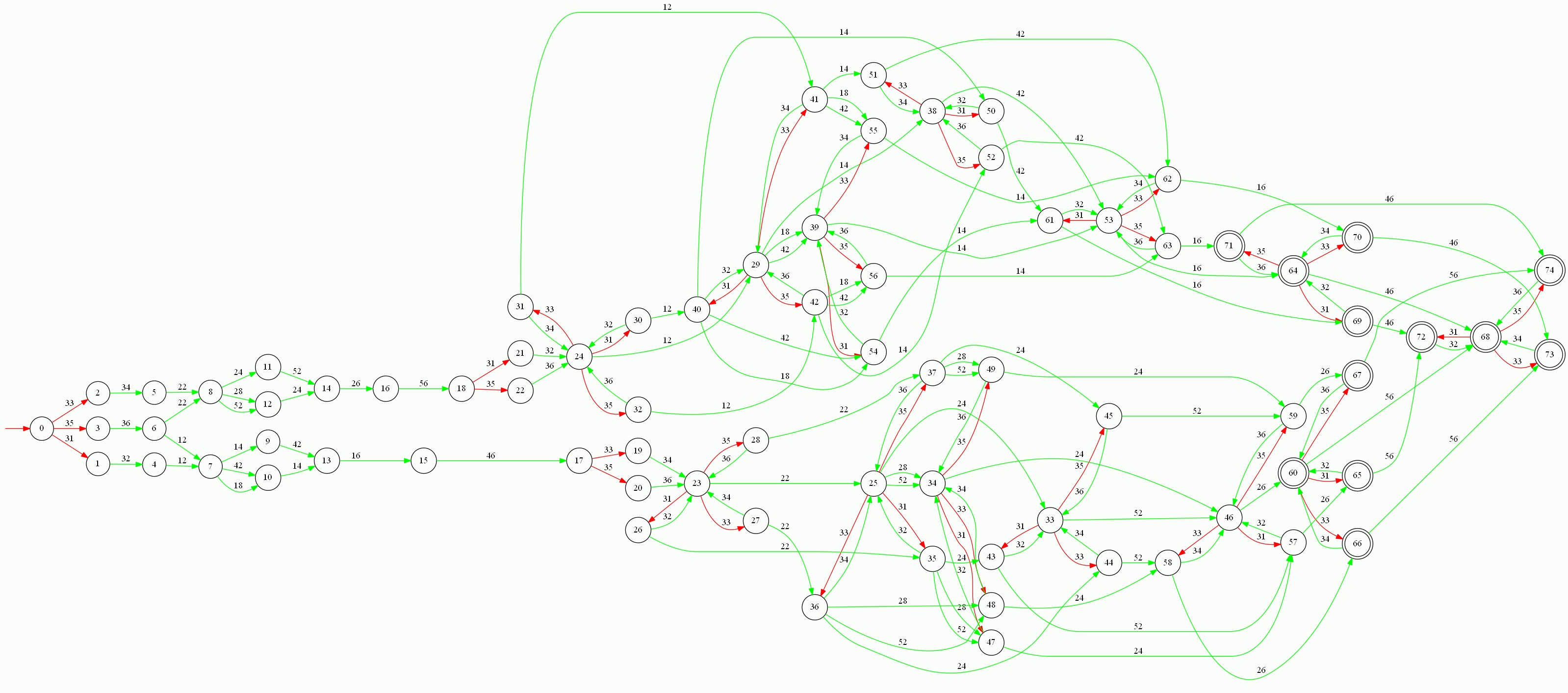}

  \caption{The synthesized supervisor $S$}

  \label{fig:result}

\end{figure*}

\begin{figure}
    \centering
    \includegraphics[scale = 0.4]{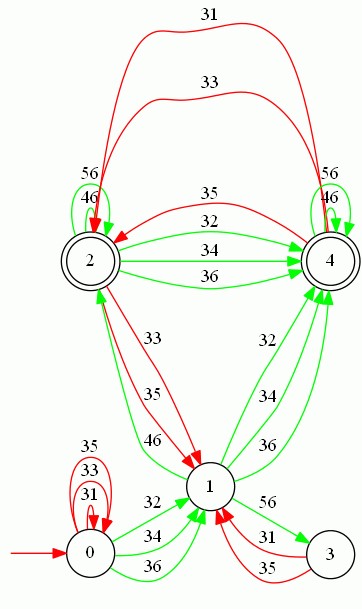}
    \caption{The synthesized supervisor after supervisor reduction}
    \label{fig:reduced}
\end{figure}

The overall transformed plant $P=OC\lVert CC\lVert CE \lVert G$ has 960 states and 3072 transitions.  The transformed specification is obtained from the transformed plant by pruning away those states where $V_1$ and $V_2$ collide.
We use TCT~\cite{feng2006tct} to synthesize a supervisor $S$ over the control constraint $(\Gamma^{in}, \Gamma^{in} \cup \Sigma_o^{out})$, by using the {\bf supscop} operation which computes the supremal normal and controllable sublanguage. The synthesized supervisor has 75 states and 174 transitions, which is shown in Fig.~\ref{fig:result}. We perform supervisor reduction in TCT by using the {\bf supreduce} operation, 
  resulting in a reduced but equivalent supervisor with 5 states and 25 transitions; the reduced supervisor  is shown in Fig.~\ref{fig:reduced}. We remark that TCT can be directly used for this example since $P$ is deterministic\footnote{It can be checked that those states with non-deterministic transitions (e.g., state 5 and state 17 of $OC$) in $OC$ cannot be reached in $P$ for this example. With this observation, Supremica~\cite{akesson2003supremica} can also be used for the synthesis, after removing those states of $OC$ with non-deterministic transitions.}. The same example has also been carried out by using the {\bf make\_supervisor} operation of SuSyNA and the same (non-reduced) supervisor is obtained. 

Intuitively, the mini-guideway example has a non-empty solution because the observation messages sent from the sensor at the junction Track2-StationB will not get lost. A supervisor may allow any  train to pass through the junction StationA-Track1 and then only needs to wait patiently until it receives an observation message from the sensor at the junction Track2-StationB; after receiving that observation message, the supervisor can release the second train. This control logic can be seen from the supervisor $S$ in Fig.~\ref{fig:result}. The reduced supervisor has self-loops of events 31, 33, 35 at the initial state, thus allowing the supervisor to issue a sequence of control messages in immediate succession. However, from the command execution automaton, only the first control message will be received, while the rest will be thrown away. 


\section{Conclusions and Future Works}
In this work, we have addressed the problem of synthesis of networked supervisors, considering bounded network delays and  also message losses in the channels. The solution methodology is to transform the networked supervisor synthesis problem to the problem of non-networked supervisor synthesis for non-deterministic plants. This approach is applicable regardless of the 
message ordering properties of the channels. It follows that we do not need to invent new properties, such as network controllability and network observability, or develop new synthesis algorithms. The effectiveness of our approach is illustrated on a mini-guideway example that is adapted from the literature, for which the supremal networked supervisor has been synthesized in the synthesis tools SuSyNA and TCT. It is not difficult to extend this approach to working with the timed discrete-event systems framework~\cite{brandin1994supervisory}, upon which the two assumptions of a) bounded consecutive losses of control messages and b) eventual observability can be removed; it is also of interest to extend this approach to the distributed control setup, the details of which will be presented elsewhere. This work also paves the way for us to consider the synthesis of covert attackers in the networked setup, which can also be transformed to the non-networked supervisor synthesis problem~\cite{Lin2020},~\cite{Lin2020J},~\cite{Lin2020JJ}. 
\bibliographystyle{elsarticle-num}
\bibliography{main}

\begin{thebibliography}{10}
\expandafter\ifx\csname url\endcsname\relax
  \def\url#1{\texttt{#1}}\fi
\expandafter\ifx\csname urlprefix\endcsname\relax\def\urlprefix{URL }\fi
\expandafter\ifx\csname href\endcsname\relax
  \def\href#1#2{#2} \def\path#1{#1}\fi

\bibitem{balemi1992supervision}
S.~Balemi, U.~Brunner, Supervision of discrete event systems with communication
  delays, in: 1992 American Control Conference, IEEE, 1992, pp. 2794--2798.

\bibitem{balemi1994input}
S.~Balemi, Input/output discrete event processes and communication delays,
  Discrete Event Dynamic Systems 4~(1) (1994) 41--85.

\bibitem{park2002robust}
S.~Park, J.~Lim, Robust and nonblocking supervisory control of nondeterministic
  discrete event systems using trajectory models, IEEE Transactions on
  Automatic Control 47~(4) (2002) 655--658.

\bibitem{park2006delay}
S.~Park, K.~Cho, Delay-robust supervisory control of discrete-event systems
  with bounded communication delays, IEEE Transactions on Automatic Control
  51~(5) (2006) 911--915.

\bibitem{park2007supervisory}
S.~Park, K.~Cho, Supervisory control of discrete event systems with
  communication delays and partial observations, Systems \& control letters
  56~(2) (2007) 106--112.

\bibitem{lin2014control}
F.~Lin, Control of networked discrete event systems: dealing with communication
  delays and losses, SIAM Journal on Control and Optimization 52~(2) (2014)
  1276--1298.

\bibitem{Shu2015}
S.~Shu, F.~Lin, Supervisor synthesis for networked discrete event systems with
  communication delays, IEEE Transactions on Automatic Control 60~(8) (2015)
  2183--2188.

\bibitem{Lin:17}
S.~Shu, F.~Lin, Deterministic networked control of discrete event systems with
  nondeterministic communication delays, IEEE Transactions on Automatic Control
  62~(1) (2017) 190--205.

\bibitem{Shu:17}
S.~Shu, F.~Lin, Predictive networked control of discrete event systems, IEEE
  Transactions on Automatic Control 62~(9) (2017) 4698--4705.

\bibitem{CDC:17}
M.~V.~S. Alves, L.~K. Carvalho, J.~C. Basilio, Supervisory control of timed
  networked discrete event systems, in: 2017 IEEE 56th Annual Conference on
  Decision and Control (CDC), 2017, pp. 4859--4865.

\bibitem{rashidinejad2018supervisory}
A.~Rashidinejad, M.~Reniers, L.~Feng, Supervisory control of timed
  discrete-event systems subject to communication delays and non-fifo
  observations, IFAC-PapersOnLine 51~(7) (2018) 456--463.

\bibitem{rashidinejad2019}
A.~Rashidinejad, M.~Reniers, M.~Fabian, Supervisory control of discrete-event
  systems in an asynchronous setting, in: 15th International Conference on
  Automation Science and Engineering, IEEE, 2019, pp. 494--501.

\bibitem{shu2014decentralized}
S.~Shu, F.~Lin, Decentralized control of networked discrete event systems with
  communication delays, Automatica 50~(8) (2014) 2108--2112.

\bibitem{tripakis2004decentralized}
S.~Tripakis, Decentralized control of discrete-event systems with bounded or
  unbounded delay communication, IEEE Transactions on Automatic Control 49~(9)
  (2004) 1489--1501.

\bibitem{park2007decentralized}
S.~Park, K.~Cho, Decentralized supervisory control of discrete event systems
  with communication delays based on conjunctive and permissive decision
  structures, Automatica 43~(4) (2007) 738--743.

\bibitem{Komenda:16}
J.~Komenda, F.~Lin, Modular supervisory control of networked discrete-event
  systems, in: Discrete Event Systems (WODES), 2016 13th International Workshop
  on, IEEE, 2016, pp. 85--90.

\bibitem{LinWang19}
F.~Lin, W.~Wang, L.~Han, S.~B., State estimation of multichannel networked
  discrete event systems, IEEE Transactions on Control of Network Systems 7~(1)
  (2019) 53--63.

\bibitem{Alves2019ACC}
M.~V. Alves, J.~C. Basilio, State estimation and detectability of networked
  discrete event systems with multi-channel communication networks, in:
  American Control Conference, 2019, pp. 5602--5607.

\bibitem{zhusupervisor}
Y.~Zhu, L.~Lin, S.~Ware, R.~Su, Supervisor synthesis for networked discrete
  event systems with communication delays and lossy channels, IEEE Conference
  on Decision and Control (2019).

\bibitem{wonham2015supervisory}
W.~M. Wonham, K.~Cai, Supervisory control of discrete-event systems, Springer,
  2018.

\bibitem{Ruochen2020}
R.~Tai, L.~Lin, Y.~Zhu, R.~Su, A new modeling framework for networked
  discrete-event systems, Automatica (under review, 2020).

\bibitem{su2010model}
R.~Su, J.~van Schuppen, J.~Rooda, Model abstraction of nondeterministic
  finite-state automata in supervisor synthesis, IEEE Transactions on Automatic
  Control 55~(11) (2010) 2527--2541.

\bibitem{Yin}
X.~Yin, S.~Lafortune, Synthesis of maximally-permissive supervisors for the
  range control problem, IEEE Transactions on Automatic Control 62~(8) (2017)
  3914--3929.

\bibitem{su2004supervisor}
R.~Su, W.~M. Wonham, Supervisor reduction for discrete-event systems, Discrete
  Event Dynamic Systems 14~(1) (2004) 31--53.

\bibitem{feng2006tct}
L.~Feng, W.~M. Wonham, Tct: A computation tool for supervisory control
  synthesis, in: 2006 8th International Workshop on Discrete Event Systems,
  IEEE, 2006, pp. 388--389.

\bibitem{su2018generalized}
URL, \href{https://www.ntu.edu.sg/home/rsu/Downloads.htm}{Susyna: Supervisor
  synthesis for non-deterministic automata}, in:
  https://www.ntu.edu.sg/home/rsu/Downloads.htm, accessed 2020.
\newline\urlprefix\url{https://www.ntu.edu.sg/home/rsu/Downloads.htm}

\bibitem{akesson2003supremica}
R.~Malik, K.~Akesson, H.~Flordal, M.~Fabian, Supremica—an efficient tool for
  large-scale discrete event systems, in: IFAC-PapersOnLine, Vol.~50, 2017, pp.
  5794 -- 5799.

\bibitem{ZMM05}
S.~Hashtrudi~Zada, M.~Moosaei, W.~M. Wonham, On computation of supremal
  controllable, normal sublanguages, Systems \& Control Letters 54 (2005)
  871--876.

\bibitem{WLLW18}
D.~Wang, L.~Lin, Z.~Li, W.~M. Wonham, State-based control of discrete-event
  systems under partial observation, IEEE Access 6 (2018) 42084--42093.

\bibitem{Lin2020T}
L.~Lin, R.~Su, A topological approach for computing supremal sublanguages for
  some language equations in supervisory control theory, Automatica (under
  review, 2020).

\bibitem{brandin1994supervisory}
B.~Brandin, W.~M. Wonham, Supervisory control of timed discrete-event systems,
  IEEE Transactions on Automatic Control 39~(2) (1994) 329--342.

\bibitem{Lin2020}
L.~Lin, Y.~Zhu, R.~Su, Synthesis of covert actuator attackers for free,
  Discrete Event Dyn Syst 30 (2020) 561–577.

\bibitem{Lin2020J}
L.~Lin, R.~Su, Synthesis of covert actuator and sensor attackers as supervisor
  synthesis, Workshop on Discrete Event Systems (in press, 2020).

\bibitem{Lin2020JJ}
L.~Lin, R.~Su, Synthesis of covert actuator and sensor attackers, Automatica
  (under review, 2020).

\end{thebibliography}








\end{document}